\documentstyle{res}
\begin{document}
\input epsf

\title{%
REHEATING AND  PREHEATING\\
 AFTER INFLATION}

\author{Lev KOFMAN \\
{\it  Institute for Astronomy,
University of Hawaii, 2680 Woodlawn Dr., Honolulu, HI 96822, USA,
kofman@kot.ifa.hawaii.edu}}

\maketitle

\section*{Abstract}

It is assumed that during inflation, all  energy
was contained in a slow-rolling inflaton field $\phi$. The
particles constituting the Universe are created due to  interactions with
the field $\phi$  coherently oscillating after inflation.
The leading channel of the particle production  would be  the non-perturbative
regime of parametric resonance, {\it preheating}. Some of the
recent developments in the theory of preheating are
 briefly reviewed. As a prototype we  use
  the model  ${\lambda\over 4}\phi^4 + {m^2 \over 2}\phi^2+
 {g^2\over 2}\phi^2\chi^2$ in an expanding universe. In different domains
of parameters the character of preheating $(\phi \to \chi)$ or
$(\phi \to \delta \phi)$
 is different, ranging
from regular Lam\'{e}  to stochastic parametric
 resonances. Fortunately,
in many important cases  simple  analytic resonant solutions are found.
 Eventually, a picture which unifies the different regimes  emerges.

\section{Creation of Particles from Inflatons}

In the inflationary scenario
 the Universe initially expands quasi-exponentially
in a vacuum-like state with   vanishing entropy and particle
number densities.
At the stage of inflation, all energy is
concentrated in a classical slowly moving inflaton field $\phi$.
Consider a simple chaotic inflation.
The fundamental
Lagrangian ${\cal L}(\phi, \chi, \psi, A_i, h_{ik}, ...)$
contains the inflaton part with the potential  $V(\phi)$
and other fields $(\phi, \chi, \psi, A_i, h_{ik}, ...)$
which give subdominant contributions to gravity.
The Friedmann equation for a scalar factor $a(t)$:
$H^2={{ 8\pi} \over {3 M_p^2}}\biggl( {1 \over 2}\dot \phi^2 +
V( \phi) \biggr)$ (with $H={\dot a / a}$)   and
the Klein-Gordon equation for $\phi(t)$:
$\ddot \phi + 3H \dot \phi + V_{, \phi}=0$
define the evolution of the background fields.
Soon
after the end of inflation,
an almost homogeneous inflaton field $\phi(t)$   coherently
oscillates with a very large  amplitude of the order of the Planck mass
 $\phi \simeq 0.1 M_p$. This scalar field can be considered as
a coherent superposition of $\phi$-quasiparticles with zero momenta,
{\it inflatons} at  rest.
 The amplitude of  oscillations  gradually
decreases not only because of the expansion of the
universe, but also because of the energy transfer to particles
created by the
oscillating field.
At this stage
we shall invoke the rest of the fundamental Lagrangian
 ${\cal L}(\phi, \chi, \psi, A_i, h_{ik}, ...)$ which
includes the other bosons  and fermions and their interactions with the
inflaton field, as well as the self-interaction of the inflaton field.
These interactions
  lead  to the  creation of many ultra-relativistic
particles from  inflatons.
In this scenario
 all the matter constituting the universe
would be created from this process of reheating.
A toy  model describing the interaction between inflatons and other
massless
Bose particles $\chi$ will be ${\cal L}=-{1 \over 2} g^2 \phi^2 \chi^2$.
Gradually, the inflaton field decays and transfers
all of its energy  non-adiabatically  to the created
particles. If the creation of particles is sufficiently slow,
   the  particles would   simultaneously
 interact with each other and come to a state of thermal equilibrium
at  the reheating temperature $T_r$.
This gradual reheating can be treated with the perturbative theory of
particle creation and thermalization.
 That  elementary theory  of reheating was
developed long  ago  with the first models of inflation.
Consider the Heisenberg representation of the quantum scalar field $\hat \chi$,
with the eigenfunctions $\chi_{k}(t)\, e^{ -i{{\bf k}}{{\bf x}}}$
 with comoving momenta ${\bf k}$.
 The temporal
part of the eigenfunction   obeys the equation
\begin{equation}
\ddot \chi_k + 3{{\dot a}\over a}\dot \chi_k + {\left(
{{\bf k}^2\over a^2}   - \xi R + g^2\phi^2 \right)} \chi_k = 0 \ ,
\label{2}
\end{equation}
with the vacuum-like initial condition $ \chi_k \simeq {e^{ -ikt} \over
 \sqrt{2k}}$
in the far past. The
coupling to the curvature $\xi R$ will not be important.
Let us seek the  solutions of Eq. (\ref{2})
 in the adiabatic WKB approximation
\begin{equation}
a^{3/2}\chi_k(t) \equiv X_k(t) =
{\alpha_k(t)\over \sqrt{2\omega}}\ e^{- i\int^t
\omega dt}
 + {\beta_k(t)\over \sqrt {2\omega}}\ e^{+ i\int^t \omega
dt} \ ,
\label{3}
\end{equation}
where the time-dependent  frequency is $\omega_k^2(t)= {{\bf k}^2 \over a^2}
+ g^2\phi^2$ plus a small correction $\sim H^2, \dot H$; initially
 $\beta_k(t)=0$. For $|\beta_k| \ll 1$, an iterative  solution\\
  $\beta_k \simeq
{\textstyle {1 \over 2}} \int\limits_{ - \infty}^{t} d t'\,{\dot \omega \over
\omega^2}\, \exp{\bigl( -
 2i \int\limits_{ - \infty}^{t'}d t'' \omega( t'')\bigr)} \sim g^2$
gives the standard result of the perturbative theory for the
particle occupation number  $n_k= \vert \beta_k \vert^2$.
This  can be interpreted as a   pair of inflatons
decaying independently  into a pair of $\chi$-particles,
$(\phi \phi \to \chi \chi)$.
The largest total decay rate in the perturbative reheating
is $\Gamma < 10^{-20} M_p$; the upper bound on the perturbative reheating
temperature is $T_r < 10^{9}$ Gev.

Recently it was found, however,  that in many
 versions of chaotic inflation reheating begins with a stage of
parametric resonance$^1$.
Indeed, in Eq.~(\ref{2}) inflatons  act not as separate
particles, but as a the coherently oscillating field $\phi(t)$.
The smallness of $g^2$ alone does
not necessarily correspond to small
$\vert \beta_k \vert$. The oscillating effective frequency $\omega_k(t)$ can
result in a broad parametric amplification of  $\chi_k$.
 In this case  the energy is rapidly
transferred from the inflaton field to other bose fields
interacting with it. This process occurs far away from thermal equilibrium, and
therefore we called it {\it preheating}.
 Because of the  non-perturbative character of the particle
 creation, the theory of preheating is rather
complicated.\\
Reheating is an intermediate stage between  inflation and the
hot radiation dominated Universe. Therefore, during or around
 reheating one can expect  initial conditions for many parameters of the
  Big Bang to be settled: the initial temperature  $T_r$,
   baryogenesis,   relic  monopoles and others  topological defects,
  primordial black holes, dark matter relics, etc.
Recent developments in the theory of reheating show that the
 stage of preheating can significantly  change the way we think
about these problems.

Among many exciting topics in preheating,   I have
selected a particular issue  for  these proceedings:
 how diverse  the character of the
parametric resonance can be in different models of $V(\phi)$,
and  how it can be  unified. This information is vital for  finding the
leading channel of the inflaton decay.
 The answers are based on a
study done in collaboration with Andrei Linde (Stanford), Alexei Starobinsky
(Moscow) and  Patrick Greene (Hawaii)$^{1,2,3,4}$.
This work was supported by NSF Grant No. AST95-29-225.

\section{Stochastic Resonance in  $ {m^2 \over 2}\phi^2+
 {g^2\over 2}\phi^2\chi^2$ Theory}

It turns out that the character of the parametric resonance
for the general equation (\ref{2})  depends on the
character of the classical oscillations $\phi(t)$, i.e., on the shape of the
effective inflaton potential $V(\phi)$.
We begin with the  quadratic
potential $V(\phi)={1 \over 2} m_{\phi} \phi^2$.
The background solution  corresponds to  sinusoidal
  oscillations
$\phi(t) \approx \Phi(t) ~
 \sin{\left( m_{\phi}t \right)}$, with the amplitude
 $ \Phi(t)= {M_p \over \sqrt{3\pi}}\cdot{1  \over m_{\phi}t}$
 decreasing as the universe expands
with the scalar factor
$ a(t) \propto t^{2/3}$.\\
For a the toy model the expansion of the universe
is  neglected
($a(t)=1$, $\dot a=0$, $\Phi=const$), the equation for the fluctuations
takes  the form  of a Mathieu equation
\begin{equation}
\ddot \chi_k +  + {\left(
 {k^2 \over m^2} + 4q  \sin^2z  \right)} \chi_k = 0 \ ,
\label{6}
\end{equation}
with a new time variable $z=mt$ and the essential
dimensionless coupling parameter $q= {{g^2 \Phi^2} \over{4 m^2}}$.
The properties of the solutions of the Mathieu equation
are well represented by its
stability/instability chart; see Fig.~1.
\begin{figure}[t]
\centering
\leavevmode\epsfysize=5.6cm \epsfbox{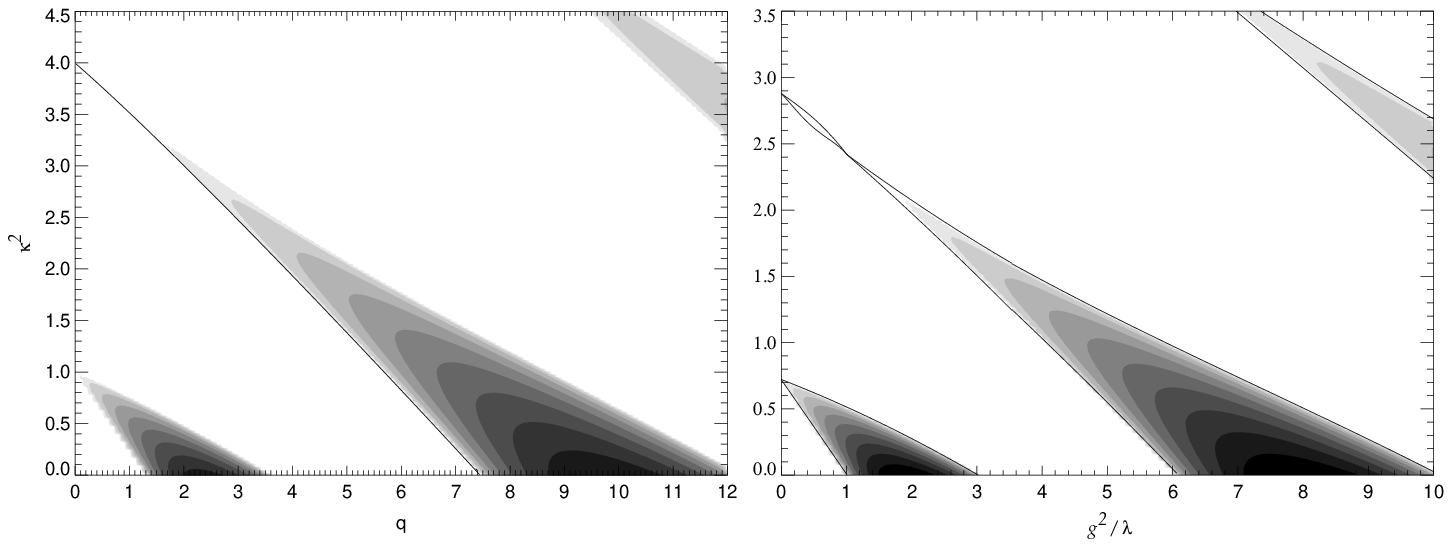}\\
\

\caption[fig8]{\label{fig1}
Stability/instability charts of the Mathieu equation (\ref{6})
 in the variables  $({k^2 \over m^2}, q)$ (left)
and of  the Lam\'{e}  equation (\ref{10})
in the variables $(\kappa^2, {g^2 \over \lambda})$ (right).
  Shaded (unshaded)
areas are regions of  instability (stability).
 For  the Lam\'{e} resonance,  the higher bands
  shrink  into nodes for    $\frac{g^2}{\lambda}={n(n+1) \over 2}$,
when simple closed form analytic solutions are found$^{3}$.
  The inflaton
 ${1 \over 4}\lambda \phi^4$  self-interaction
in $O(N)$ theory  corresponds to $n=1$  for $N=1$ and $n=3$ for  $N > 1$.
}
\end{figure}
There is an
 exponential instability $\chi_k \propto \exp (\mu_k z)$
 within the set of resonance bands.
This instability corresponds to an  exponential growth in the occupation
number of created particles
$n_{k}(t) \propto \exp (2\mu_k z)$.
Since ${m \over M_p} \simeq 10^{-6}$,
 it is expected that $q \simeq 10^{10} g^2 \gg 1$, the parametric resonance
is broad. The simple standard methods for finding the characteristic
exponent $\mu_k$ in the narrow resonance regime $q \ll 1$
 do not work here.
However, we made the following observation$^2$: for large $q$ the eigenfunction
$\chi_k(t)$ is changing adiabatically  between
the moments $t_j$, $j=1,2,3, ...$, where
the inflaton field is equal to zero
$\phi(t_j)=0$. The
 non-adiabatic changes of $\chi_k(t)$ occur only
 in the vicinity of $t_j$.
Therefore, we expect that the semiclassical solution (\ref{3})
 is valid everywhere but around
 $t_j$.  Let the wave $\chi_k(t)$
have the form of the WKB solution with the pair of coefficients
$(\alpha_k^{j}, \beta_k^{j})$
before the scattering at the point $t_j$;
and the pair $(\alpha_k^{j+1}, \beta_k^{j+1})$ after the scattering at
 $t_j$.
The interaction term  around all the points $t_j$
is parabolic
 $g^2\phi^2(t) \approx g^2\Phi^2m^2 (t-t_j)^2$.
 Therefore, the  outgoing amplitudes
 $(\alpha_k^{j+1}, \beta_k^{j+1}) $ can be expressed through
 the incoming amplitudes
$(\alpha _k^{j}, \beta_k^{j})$ with help of the well-known
 reflection
 $R_k$ and transmission $D_k$
amplitudes for scattering from a parabolic potential at
 $t_j$:
\begin{equation}\label{7}
\pmatrix{\alpha_k^{j+1} \cr \beta_k^{j+1} \cr } =
\pmatrix{  e^{i\varphi_k}  \cosh \lambda       &
         i e^{ +2i\theta_k^{j}} \sinh \lambda   \cr
        -i e^{ -2i\theta_k^{j}} \sinh \lambda        &
           e^{-i\varphi_k} \cosh \lambda         \cr}
\pmatrix{\alpha_k^{j} \cr \beta_k^{j} \cr}
 \ , \,\, \sinh \lambda=e^{-{\pi \over 2} \kappa^2} \ ,
\end{equation}
where
$\varphi_k= \arg
 \Gamma \left({1+i\kappa^2 \over 2}\right)+{\kappa^2\over 2}\left(1+
\ln{2\over\kappa^2}\right)$,
 the phase
accumulated by the moment ${t_j}$ is
$\theta_k^{j}=\int\limits_0^{t_j} dt~ \omega(t)$,
and a scaled momentum $\kappa={k \over \sqrt{g m \Phi}}$.

The growth index $\mu_k$ is defined by the formula
$n^{j+1}_k =n^{j}_k \exp( 2\pi \mu_k^{j})$.
IFrom (\ref{7}), one finds
 the characteristic exponent
in the large occupation number limit
\begin{equation}
\mu_k^{j}={1 \over 2\pi} \ln \left(
1 +2 e^{-\pi \kappa^2} -
2\sin \theta_{tot}^{j}~
e^{-{{\pi \over2} \kappa^2}}~\sqrt{1+ e^{-\pi \kappa^2}}
\right) \ .
\label{8}
\end{equation}
For the Mathieu Eq.~(\ref{6})
there is a simple solution of the
matrix equation (\ref{7}) for an arbitrary $j$, which allows us  to
find the resonance band
$| \tan (\theta_k- \varphi_k)| \leq e^{-{{\pi \over 2} \kappa^2}}$
and to fix the phase $\theta_{tot}^{j}$,
$\cos {(\theta_{k,tot})}=
\sqrt{1+e^{\pi \kappa^2}}\sin {(\theta_k-\varphi_k)}$.
The crucial expression  here is
the phase $\theta_k$
 accumulating between two successive zeros of $\phi(t)$,
$\theta_k =\int^{\pi \over m}dt \sqrt{k^2+ g^2 \phi^2(t)}$,
so that $\theta_k - \varphi_k \approx 4\sqrt{q}
+ {k^2\over 8\sqrt{q} m^2} \bigl(  \ln  q +9.474 \bigr)
 \approx {{2 g\Phi }\over m} + O(\kappa^2)$.
\begin{figure}[t]
\centering
\leavevmode\epsfysize=5.2cm \epsfbox{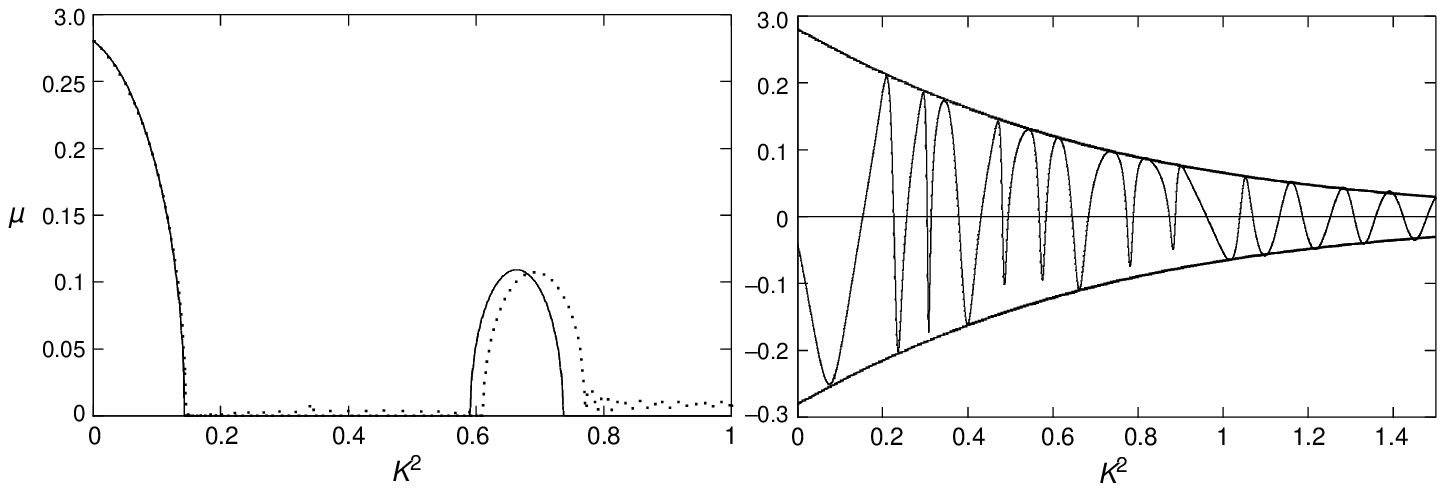}\\
\

\caption[fig8]{\label{fig8}
Left: the characteristic exponent $\mu_k$ of the
Mathieu equation (\ref{6}) as a function
of $k^2$ (in units of $\sqrt{g m \Phi}$)
for   $q \approx 10^4  $
(dotted curve). Two instability
bands are shown. The solid curve
 is derived analytically
with a simple (improvable) approximation$^2$.
Right: the characteristic exponent $\mu^j_k$ of the
mode Eq. (\ref{2}) in an expanding universe for $j=10$.
 $\mu^j_k$ changes dramatically with $j$ but within the
  envelope (bold curve) which is obtained from Eq. (\ref{8}) by taking there
$\sin \hat \theta = \pm 1$.
 Positive and negative occurrences of
$\mu_k$  for $\kappa < \pi^{-1}$   appear in the proportion $3:1$.
 }
\end{figure}
The function $\mu_k$ derived with the formulas (\ref{8})
 is independent of $j$ and
is plotted in  
Fig.~2 (left panel), which is a slice of the Mathieu stability/instability
chart  for given (large) $q$.
That transparently
shows the presence of a sequence of stability/instability bands
as a function of $k$.
Typical half-width of a resonance band is $k^2 \sim 0.1 g m \Phi$.
The  effect of parametric resonance
crucially depends on the interference of the wave functions, i.e.,
the phase correlation/anticorrelation between successive scatterings
at the parabolic potentials.
The maximal value of $\mu$ is reached for the
positive interference when $\sin \theta_{tot} =-1$ and is equal
to
 $\mu={1 \over \pi} \ln \left(1 +\sqrt{2}\right) \approx 0.28$.
The typical value of $\mu$ corresponds to $\sin \theta_{tot} =0$ and is
equal to
$\mu={1 \over 2\pi} \ln 3 \approx 0.175$.
For the negative
 interference when $\sin \theta_{tot} =1$ there is no resonance.
The angle $\theta_{tot}$ is mainly controlled by the
 phase $\theta_k^{j}$.

Let us consider now  an expanding universe. At first glance,
the resonant solution here is more
complicated than the Mathieu equation, which for $q \gg 1$
has only a  heuristic meaning for our problem.
The most important change is the time-dependence of the
  parameter $q={{ g^2 \Phi^2} \over 4 m^2}$:
 $q  \propto j^{-1} $;
$2j$ is the number of inflaton oscillations.
 For the broad resonance case
this parameter significantly varies within
a few inflaton oscillations; hence, the concept of the static
 stability/instability chart and the results for
 the Mathieu equation cannot be utilized here.
Surprisingly, the most interesting case, when the parameter $q$
is large {\it and} time-varying, can also
 be treated analytically
by the method of  successive parabolic scatterings.
Indeed, the matrix mapping (\ref{7}) for the
$\alpha_k^j$ and $\beta_k^j$  is also valid
in the case of an expanded universe.
The phase accumulating between two successive
zeros of the inflaton field is now
$\theta_k^j \approx {{g M_p} \over {5 m j}}+ O( \kappa^2)$.
For large initial values of $q$,
 the phase rotation
  between successive scatterings
$\delta \theta_k  \simeq    {{ \sqrt{q}} \over 8j^2}$
 is much larger than $\pi$
 for all relevant
$k$. Therefore,  the phase $\theta_{tot}$ in formula
(\ref{8})  can be considered to be random numbers.
Since there is no memory of the phases,
 each mapping  (\ref{7})    can be considered as
independent  of the previous ones.
As a result the number of particles $n_k^{j}$ and
the   exponent $\mu_k$ are random variables.
The functional form of $\mu_k$ for the stochastic resonance is
different from that for the broad parametric resonance,
in (\ref{8}) one has substitute a random phase
 $\hat \theta$ in the interval $[0, 2\pi]$ instead of
$\theta_{tot}$.
An example of a stochastic $\mu_k$
 is plotted in Fig.~2.\\
Contrary to the regular resonance,
the stochastic resonance is much broader, there are no
distinguished stability/instability bands, and for certain values of momenta
the function $\mu_k^j$ is negative. During the stochastic resonance regime,
this function changes dramatically with every half   period of
 the inflaton oscillations.    Fig.~\ref{fig8}
shows that it is incorrect to use the structure of the resonance bands of the
static Mathieu equation for investigation of the stage of stochastic resonance,
unless the parameter $q$ drops to a value $\sim O(1)$.

\section{Structure of the Resonance in  ${\lambda\over 4}\phi^4 +
 {g^2\over 2}\phi^2\chi^2$ Theory}

In the theory with  the potential
 $V(\phi)={1 \over 4}\lambda \phi^4$
it is  convenient
to use the conformal field $\varphi=a \phi$ and
 conformal  time variable
 $\tau= \sqrt{\lambda   \tilde\varphi^2 }\int {dt\over a(t)}$.
Then (without decay) the amplitude of  $\varphi$
   is constant,  $\tilde\varphi$.
The background solution  is given by an elliptic function
$\varphi(\tau) \approx   \tilde\varphi  ~
  cn \left( \tau, { 1 \over \sqrt{2}}\right)$,
  $a(\tau)=\sqrt{2 \pi \over 3 }{\tilde\varphi\over M_p}\tau$.
The Eq.~(\ref{2})   for quantum fluctuations $\chi_k$
 can be simplified in this theory.
Indeed, the physical momentum,  ${\bf p} = {{\bf k} \over a(t)}$
 is redshifted
in the same manner as the background field
amplitude.
The conformal transformation of the mode function
$ X_k(t)= a(t)\chi_k(t) $ allows one to
 rewrite  Eq.~(\ref{2}) as
\begin{equation}
X_k''  +  {\left(\kappa^2  +
 {g^2\over \lambda}  cn^2 \Bigl(\tau,  { 1 \over \sqrt{2}}\Bigr)
 \right)} X_k  = 0 \ ,
\label{10}
\end{equation}
where $\kappa^2={ k^2 \over  \lambda   \tilde\varphi^2}$.
The equation for fluctuations does not depend on the
expansion of the universe and is completely reduced to the similar
problem in Minkowski space-time.
This is a special feature of the
conformal  theory
 ${1 \over 4} \lambda \phi^4  + {1 \over 2} g^2 \phi^2 \chi^2 $.
The mode equation  (\ref{10})
belongs to the class of   Lam\'{e} equations.
Since the coefficients of the  Lam\'{e} equation (\ref{10})
are time-independent, we can numerically construct the
 stability/instability chart in the variables
$\left(\kappa^2, {{g^2} \over \lambda} \right)$, plotted in Fig. 1.
The combination of parameters  $g^2/\lambda$   ultimately
defines the structure of the parametric resonance in this theory.
This means that the condition of  a  broad parametric resonance
does not require a large initial amplitude of the inflaton field,
 as for the quadratic potential.
The strength of the resonance
depends non-monotonically
on this parameter.

\section{Unified Picture for ${\lambda\over 4}\phi^4 + {m^2 \over 2}\phi^2+
 {g^2\over 2}\phi^2\chi^2$ Theory}

Preheating in  the
theory ${m^2\over 2}\phi^2 + {g^2\over 2} \phi^2\chi^2$
is efficient only if    $g\Phi \gg m$. The inflaton amplitude
$\Phi$
is extremely large  immediately after inflation, $\phi \simeq 10^{-1} M_p$, and
later it decreases as $\Phi \sim {M_p\over 3mt}$.
The parameter   $q$, which controls
the resonance, rapidly changes. As a
result, the broad parametric resonance regime in this theory is a stochastic
process.
In the theory   ${\lambda\over 4}\phi^4 + {g^2\over 2}
\phi^2\chi^2$ the inflaton field $\phi$ also decreases in an expanding
universe, but it does not make the resonance stochastic because all
variables  scale in the same way as $\phi$ due to the conformal
invariance of the model.
The parameter which controls the strength of the resonance is
 ${g^2 \over \lambda}$.
What is the relation between these two theories?
Neither of these two theories is completely general. In the theory of
the massive scalar field one may expect  terms $  {\lambda\over 4}\phi^4$
to appear because of radiative corrections. On the other hand, in many
realistic theories the effective potential is quadratic with respect to $\phi$
near the minimum of the effective potential.

Let us consider the theory ${m^2\over 2}\phi^2 +
{\lambda\over 4}\phi^4 + {g^2\over 2} \phi^2\chi^2$ which embraces both limits.
\begin{figure}[t]
\centering
\leavevmode\epsfysize=5.4cm \epsfbox{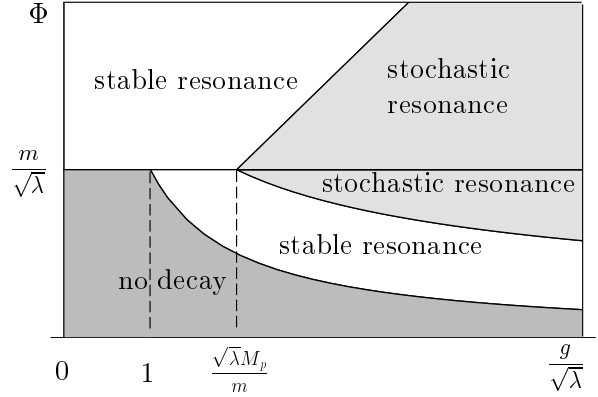}\\
\

\caption[ResRanges]{\label{ResRanges} Schematic representation of different
regimes which are possible in the theory ${m^2\over 2}\phi^2 + {\lambda\over
4}\phi^4 + {g^2\over 2} \phi^2\chi^2$
 for various relations between $g^2$ and $\lambda$ in an expanding
universe. }
\end{figure}
 One may expect that for
$\phi \gg { m\over\sqrt\lambda}$ the mass term ${m^2\over 2}\phi^2$
does not affect the frequency of oscillations of the inflaton field $\phi$, and
the parametric resonance in this theory occurs in
the same way as in the conformal theory,
whereas for  $\phi \ll  {m\over\sqrt\lambda}$ the resonance
develops as in the massive   theory.
Therefore, a naive thought would be  that as the amplitude $\Phi$
decreases from $10^{-1}M_p$ to ${m\over\sqrt\lambda}$,
the parametric  resonance  can be described with  the
Lam\'{e} stability/instability
chart. It turns out that the actual criterion for the regular, stable
 resonance
is that the phase of the $\chi_k$-wave between two zeros
of inflaton oscillations, $\theta_k=\int^{T \over 2} d\tau \sqrt{\kappa^2 +
{g^2 \over \lambda} \varphi^2(\tau)}$ is not rotating.
The phase rotation  within an inflaton  half-period is
$ \delta \theta_k \simeq {g  \pi^2 m^2 H \over 4\lambda^2 \Phi^3 } $.
This means that despite the subdominant contribution of the  mass term,
 for large ${g^2\over \lambda}$ its presence  can modify
the nature of the  Lam\'{e} resonance by turning it into
 stochastic process if
$\Phi \leq {g\over\sqrt\lambda}\, { \pi^2 m^2  \over 3\lambda  M_p}$.
Different regimes of parametric resonance  are shown in Fig.~\ref{ResRanges}.
 Immediately after inflation the amplitude
$\Phi$  is greater than
${m\over\sqrt\lambda}$. If ${g\over \sqrt\lambda} \leq {\sqrt\lambda
M_p\over   m}$, the $\chi$-particles are produced in the  regular
resonance regime until  $\Phi(t)$ decreases to ${m\over
\sqrt\lambda}$, after which the resonance occurs as in the theory ${m^2\over
2}\phi^2   + {g^2\over 2} \phi^2\chi^2$. The resonance never
becomes stochastic.
If  ${g\over \sqrt\lambda} \geq {\sqrt\lambda M_p\over   m}$, the resonance
originally develops as in the conformal  theory,
 but with a decrease of $\Phi(t)$ the
resonance becomes stochastic. Again, for
$\Phi(t) \leq {m\over \sqrt\lambda}$
 the resonance occurs as in the theory
${m^2\over 2}\phi^2   + {g^2\over 2} \phi^2\chi^2$.
There is a transient regime  of intermittancy between regular (stable) and
 stochastic resonances.
 In all cases the resonance
eventually disappears when the field $\Phi(t)$ becomes sufficiently small
and is replaced by the perturbative decay.
Perturbative  reheating
can be complete only if there is  symmetry breaking in the theory,
or additional interactions with fermions$^{2,4}$.

\vspace{1pc}

\re
1.   Kofman, L.,  Linde, A. \& Starobinsky, A. 1994,
 {\it  Phys.\ Rev.\ Lett.} {\bf 73}, 3195.
\re
2.   Kofman, L.,  Linde, A. \& Starobinsky, A. 1997,
{\it Phys.\ Rev.} {\bf D56},  3258.
\re
3. Greene, P.,  Kofman, L.,  Linde, A., \& Starobinsky, A. 1997, {\it Phys.\ Rev.}{\bf D56},  6175.
\re
4.  Kofman, L. In:
{\it Relativistic Astrophysics: Novikov's 60th Birthday. Conference}
  Cambridge Univer. Press 1997, p.133.

\vspace{1pc}

\newpage
%%%%%%%%%%%%%%%%%%%%%%%%%%%%%%%%%%%%%%
%       Please fill out items listed below. See %%% EXAMPLE %%%
%
%%%%%%%%%%%%%%%%%%%%%%%%%%%%%%%%%%%%%%

\chapter*{ Entry Form for the Proceedings }

\section{Title of the Paper}
%%%%%%%%%%%%%%%%% Enter the title of your paper.

{\Large\bf %

REHEATING AND PREHEATING\\
AFTER INFLATION
}

\section{Author}

\newcounter{author}
\begin{list}%
{}{\usecounter{author}}

%%%%%%%%%%%%%%%%% This item unit is just for one author.
\item %
\begin{itemize}
\item Full Name:                Lev Kofman %%% EXAMPLE %%%
\item First Name:               Lev %%% EXAMPLE %%%
\item Middle Name:               %%% EXAMPLE %%%
\item Surname:                  Kofman %%% EXAMPLE %%%
\item Initialized Name:         L. Kofman
 %%% EXAMPLE %%%
\item Affiliation:              IfA, Honolulu, Hawaii, USA %%% EXAMPLE %%%
\item E-Mail:                   kofman@kot.ifa.hawaii.edu %%% EXAMPLE %%%
\item Ship the Proceedings to:  Honolulu, Hawaii, USA %%% EXAMPLE %%%
\end{itemize}

\end{list}

\end{document}